\documentclass[10pt]{article}

\usepackage{amsmath,amsthm,amssymb,amsfonts}

\usepackage[a4paper]{geometry}

\usepackage{a4wide}

\usepackage{graphicx}

\usepackage{subfigure}

\usepackage{url}

\usepackage{hyperref}

\usepackage{listings}
\usepackage{color}
\usepackage[dvipsnames]{xcolor}

\usepackage[margin=10pt,font=small,labelfont=bf, labelsep=endash]{caption}

\newcommand{\defeq}{\overset{\text{def}}{=}}

\usepackage{authblk}

\title{Multi-Agent Shape Control with Optimal Transport}
\author[1]{Alex Tong Lin}
\author[1]{Stanley J. Osher}
\affil[1]{University of California, Los Angeles (UCLA)}

\date{\today}

\begin{document}
	
\maketitle

\begin{abstract}
	We introduce a method called MASCOT (Multi-Agent Shape Control with Optimal Transport) to compute optimal control solutions of agents with shape/formation/density constraints. For example, we might want to apply shape constraints on the agents -- perhaps we desire the agents to hold a particular shape along the path, or we want agents to spread out in order to minimize collisions. We might also want a proportion of agents to move to one destination, while the other agents move to another, and to do this in the optimal way, i.e. the source-destination assignments should be optimal. In order to achieve this, we utilize the Earth Mover's Distance from Optimal Transport to distribute the agents into their proper positions so that certain shapes can be satisfied. This cost is both introduced in the terminal cost and in the running cost of the optimal control problem.
\end{abstract}

\section{Introduction}

	Optimal control seeks to find the best policy for an agent that optimizes a certain criterion. This general formulation allows optimal control theory to be applied in numerous areas such as robotics, finance, aeronautics, and many other fields. Inherently, optimal control optimizes the control of a single agent, but in recent years, extending optimal control problems to the realm of multi-agents has been a popular trend. Indeed, there are numerous cases where we want to model not just a single agent, but many, e.g. a fleet of drones.
	
	Here we introduce MASCOT: Multi-Agent Shape Control with Optimal Transport, a method to compute solutions to multi-agent optimal control problems that involve shape, formation, or density constraints among the agents. These constraints can be formulated in the running cost of the agents, or as a terminal cost, or even both.
	
	We first introduce the reader to optimal control and its multi-agent version. We then review the idea of optimal transport and Earth Mover's Distance. Finally, we demonstrate the method on some examples.

\section{Related Works}
	In terms of multi-agent shape/formation constraints, as noted in \cite{xiao2009finite}, some methods rely on the use of leaders \cite{egerstedt2001formation, egerstedt2001control, leonard2001virtual, shao2007leader, 1656406} to help agents make formations. Others are more behavior-based \cite{arkin1998behavior, balch1998behavior, lumelsky1997decentralized}. And there are some that rely on a virtual-structure approach \cite{lewis1997high}. Methods based on consensus of agents has also been considered in \cite{lafferriere2005decentralized, ren2006consensus, xiao2009finite, 7974899}. There is also work based on potential functions \cite{zavlanos2007sensor, hengster2010multi}. There are also many algorithms for solving multi-agent optimal control problems based on different assumptions on how the multi-agents interact: \cite{movric2013cooperative, onken2021neural, foderaro2014distributed, lin2018splitting}.
	
	In terms of using the assignment problem, \cite{Kwok2002-ch} used various assignment algorithms to match agents with destinations so as to improve surveillance. In \cite{macdonald2011multi}, they experiment with solving the assignment problem in order make agent formations as the end goal, but do not place it into an optimal control framework. Recently, in \cite{9672686} they apply discrete optimal transport in a capability-aware fashion to assign $N$ agents to $N$ moving targets, although they do not consider collisions. A survey of multi-agent formation control is provided in \cite{OH2015424}.
	
	Further, related to the field of multi-agent optimal control are the field of Mean Field Control and Mean Field Games \cite{doi:10.1073/pnas.2024713118, doi:10.1073/pnas.1922204117, lauriere2022learning}.

\section{Background}

	We provide background on optimal control, multi-agent optimal control, and optimal transport.

\subsection{Optimal Control}
\label{sec:background-optimal-control}

	Optimal control seeks to find a control law that best optimizes a cost or payoff criterion. Here we will stick with convention of minimizing a cost. 

	Given an initial point $x\in\mathbb{R}^n$ and an initial time $t\in [0, T]$, the system will follow the dynamics:
	\begin{equation*}
		\left\{
		\begin{aligned}
			\dot{\textbf{x}}(s) &= \textbf{f}(\textbf{x}(s), \textbf{u}(s)), &\quad t < s < T \\
			\textbf{x}(t) &= x
		\end{aligned}
		\right.
	\end{equation*}
	where $x$, $\textbf{x}(s) \in \mathbb{R}^n$ for all $s\in(0, T)$, $\textbf{f}: (\mathbb{R}^n \times U) \rightarrow \mathbb{R}^n$, and $U\subseteq \mathbb{R}^m$. We call $\textbf{x}$ the \emph{state}, and $\textbf{u}$ the \emph{control}. Then we want to minimize the functional $J_{x, t}:\mathcal{U} \rightarrow \mathbb{R}$ where,
		\begin{equation*}
			J_{x,t}[\textbf{u}] \defeq g(\textbf{x}(T)) + \int_t^T L(\textbf{x}(s), \textbf{u}(s))\,ds
		\end{equation*}
	and where $\mathcal{U} \defeq \{\textbf{u} : \textbf{u}: (0, T) \rightarrow U\}$ is called the \emph{admissable control} set, $g:\mathbb{R}^n \rightarrow \mathbb{R}$ is the \emph{terminal cost} and $L: (\mathbb{R}^n \times U) \rightarrow \mathbb{R}$ is the \emph{running cost}. So optimal control seeks to find $\textbf{u}\in \mathcal{U}$ that minimizes $J_{x, t}$.

\subsection{Multi-Agent Optimal Control}
\label{sec:background-multi-agent-optimal-control}

	There are many formulations of multi-agent optimal control, based on whether the control is centralized or decentralized, or whether the agents communicate or not, and many other factors. In this work, we consider a simple extension of optimal control with a centralized controller and a finite number of agents. Later, the control can perhaps be decentralized with an imitation learning algorithm as demonstrated in \cite{lin2022decentralized}.

	In this work, we consider the following multi-agent control problem where the dynamics modeling the $N$ agents are:
		\begin{equation*}
			\left\{
			\begin{aligned}
				\dot{\textbf{x}}^{(i)}(s) &= \textbf{f}(\textbf{x}^{(i)}(s), \textbf{u}^{(i)}(s)), &\quad t < s < T, \quad 1 \le i \le N \\
				\textbf{x}^{(i)}(t) &= \textbf{x}^{(i)}_0
			\end{aligned}
			\right.
		\end{equation*}
	and they want to collectively minimize the following cost functional:
		\begin{equation*}
			J[\{\textbf{u}^{(i)}\}_{i=1}^N] \defeq g(\{\textbf{x}^{(i)}(T)\}_{i=1}^N) + \int_t^T L(\{\textbf{x}^{(i)}(s)\}_{i=1}^N, \{\textbf{u}^{(i)}(s)\}_{i=1}^N)\,ds.
		\end{equation*}

	We note that technically, if we stack the states $\{\textbf{x}^{(i)}\}$ into one concatenated vector, and we stack the controls $\{\textbf{u}^{(i)}\}$ also into one concatenated vector, then this can viewed as a single-agent control problem. This would model the realistic situation where a fleet of drones are being controlled by a centralized controller, for example.

\subsection{Earth Mover's Distance and Optimal Transport}
\label{sec:background-emd}

	The problem of Optimal Transport seeks to find a transportation plan between two probability distributions that is optimal. The cost of the plan also provides a distance metric between the probability distributions, called \emph{Earth Mover's Distance} (also called the \emph{Wasserstein distance}).
	
	To aid explanation and because this is the most relevant case to us, we restrict ourselves to discrete distributions: Suppose we are given two sets of points: $\{\textbf{x}^{(i)}\}_{i=1}^N$ and $\{\textbf{z}^{(i)}\}_{i=1}^M$, and we weight them with distributions $\textbf{a} = (a_1, \ldots, a_N)\in\mathbb{R}^N$ and $\textbf{b} = (b_1, \ldots, b_M)\in\mathbb{R}^M$. So $\textbf{x}^{(i)}$ has weight $a_i$ for example. Then we can define a cost matrix,
		\begin{equation*}
			C = (c_{ij}) \in \mathbb{R}^{N \times M}, \quad c_{ij} = \|a_i - b_j\|^2, \quad 1 \le i \le N, 1 \le j \le M. 
		\end{equation*}
	Then optimal transport seeks to find a transportation plan $\pi = (\pi_{ij}) \in \mathbb{R}^{N \times M}$ that minimizes the following,
		\begin{equation*}
			\text{EMD}_C(\textbf{a}, \textbf{b}) = \min_{\pi} \left\{ \sum_{ij} \pi_{ij} c_{ij} : \sum_{j} \pi_{ij} = a_i, \text{ for all $i$}, \sum_{i} \pi_{ij} = b_j, \text{ for all $j$}\right\}
		\end{equation*}
	and value of the minimum is called the \emph{Earth Mover's Distance}. The argmin is the transportation plan $\pi$.
	
	In the special case where $N = M$ and $\textbf{a}$ and $\textbf{b}$ are the uniform distribution (i.e. $a_i = b_i = 1/N$ for all $1 \le i \le N$), then we can just set $\textbf{a} = (1, \ldots, 1)$ and $\textbf{b} = (1, \ldots, 1)$. Then our problem simply becomes an assignment problem. Although, optimal transport is actually much more general and we take advantage of this: we can demand distributional preferences. For example, we may want a proportion of agents to cover one destination and the rest to cover another destination. In this case, we may have $N$ agents, but only $2$ destination points, which significantly reduces the computation cost of the cost matrix from $O(N^2)$ (using the assignment algorithm), to $O(2N) = O(N)$ (using optimal transport).
	
	Related to the computation of the Earth Mover's Distance, is the computation of a regularized Earth Mover's Distance utilizing Sinkhorn iteration \cite{NIPS2013_af21d0c9} and variants therein, which are able to achieve faster computational times, sometimes even achieving near-linear time complexity \cite{NIPS2017_491442df}.

\section{Methods}

	In this section, we provide details on how to implement our method. We first explain the case of controlling the multi-agent shape/density at terminal time. Then we explain how to implement the method for the running cost.

\subsection{Shape control at terminal time}
	
	As noted in Section~\ref{sec:background-multi-agent-optimal-control}, the agents' dynamics are,
		\begin{equation*}
			\left\{
			\begin{aligned}
				\dot{\textbf{x}}^{(i)}(s) &= \textbf{f}(\textbf{x}^{(i)}(s), \textbf{u}^{(i)}(s)), &\quad t < s < T, \quad 1 \le i \le N \\
				\textbf{x}^{(i)}(t) &= x^{(i)}
			\end{aligned}
			\right.
		\end{equation*}
	and they want to minimize the cost-functional,
		\begin{equation*}
			J[\{\textbf{u}^{(i)}\}_{i=1}^N] = g(\{\textbf{x}^{(i)}(T)\}_{i=1}^N) + \int_t^T L(\{\textbf{x}^{(i)}(s)\}_{i=1}^N, \{\textbf{u}^{(i)}(s)\}_{i=1}^N)\,ds.
		\end{equation*}
	Suppose we want the agents to satisfy the shape/density of the reference points $\{\textbf{z}^{(j)}\}_{j=1}^M$. We set the cost matrix to be,
		\begin{equation*}
			C = (c_{ij}), \quad c_{ij} = d(\textbf{x}^{(i)}(T), \textbf{z}^{(j)})
		\end{equation*}
	where $d$ is the cost between $\textbf{x}^{(i)}$ and $\textbf{z}^{(i)}$. For example, $d$ can be the squared Euclidean distance. Then we choose a distribution on the $\{\textbf{x}^{(i)}(T)\}_{i=1}^N$, which we call $\textbf{a} = (a_1, \ldots, a_N)$. We also choose a distribution on the $\{\textbf{z}^{(j)}\}_{j=1}^M$, which we call $\textbf{b} = (b_1, \ldots, b_M)$. Then we can add to the terminal cost,
		\begin{equation*}
			g_{\text{shape}}(\{\textbf{x}^{(i)}(T)\}_{i=1}^N) = \sum_{ij} \pi^*_{ij}c_{ij}
		\end{equation*}
	where $\pi^*$ is the optimal transport plan of the Earth Mover's Distance as presented in Section~\ref{sec:background-emd}, with cost matrix $C = (c_{ij})$. The Earth Mover's Distance and the transportation plan can be computed using standard linear programming, or one can use Sinkhorn iteration \cite{NIPS2013_af21d0c9} to approximate the distance and transportation plan.
	
	Intuitively, the transportation plan $\pi$ weights the cost matrix $C$, so that agents can then use this information to find optimal assignments to the proper reference points.

\subsection{Shape control in the running cost}

	In order to apply shape/density constraints along the trajectory of the agents, we first subtract the mean from the agents and the reference points:
		\begin{equation*}
			\textbf{y}^{(i)}(t) = \textbf{x}^{(i)}(t) - \bar{\textbf{x}}(t), \quad \text{and} \quad \textbf{w}^{(j)}(t) = \textbf{z}^{(j)} - \bar{\textbf{z}},
		\end{equation*}	
	where $\bar{\textbf{x}}(t) = \frac{1}{N} \sum_{i=1}^N \textbf{x}^{(i)}(t)$, and $\bar{\textbf{z}}(t) = \frac{1}{M} \sum_{j=1}^M \textbf{z}^{(j)}(t)$ are the mean. Then letting $\textbf{a}$ and $\textbf{b}$ be the weight distribution of the $\textbf{y}$ and $\textbf{w}$, as before, then shape constraints can be applied with,
		\begin{equation*}
			L_{\text{shape}}(\{\textbf{x}^{(i)}(t)\}_{i=1}^N) = \sum_{ij} \pi^*_{ij}(t)c_{ij}(t)
		\end{equation*}
	with cost matrix $C(t)=(c_{ij}(t))$ and optimal transport plan $\pi^*(t)$ as before, but now with $\textbf{y}$ and $\textbf{w}$, and depending on time.
	
	Subtracting the mean allows the agents to become agnostic to the actual position of the reference points, and now the agents can focus on the relative positioning amongst each other in order satisfy the shape constraint. This also allows the optimal control problem itself to find optimal paths for the agents, because otherwise the actual positions of the reference points would interfere. One can even try to normalize the agents and reference points using bounding boxes, so agent formations are agnostic to the actual diameter of the reference points.

\subsection{Classical Direct Shooting for Optimal Control}
\label{sec:classical-direct-shooting-for-optimal-control}

	Our approach to inducing shape constraints on the agents is generally agnostic to the method used to compute optimal control problems. Thus in our numerical demonstration, we employ a very simple and straightforward method to compute optimal control solutions -- the direct shooting method.
	
	We first discretize the time domain:
		\begin{equation*}
			0 = t_0 < t_1 < \cdots < t_S = T.
		\end{equation*}
	For convenience, we let the discretization be uniform with mesh size $\Delta t$. Then denoting  $\textbf{x}^{(i)}_s \defeq \textbf{x}^{(i)}(t_s)$ and similarly with $\textbf{u}^{(i)}_s$, if we use forward Euler approximations for the agents' dynamics, then we have,
		\begin{equation}\label{eq:agent-dynamics}
			\left\{
			\begin{aligned}
				\textbf{x}^{(i)}_{s+1} &= \textbf{x}^{(i)}_s + \Delta t\, \textbf{f}(\textbf{x}^{(i)}_s, \textbf{u}^{(i)}_s), &\quad t < s < T, \quad 1 \le i \le N \\
				\textbf{x}^{(i)}_0 &= \textbf{x}^{(i)}(0)
			\end{aligned}
			\right.
		\end{equation}
	Our cost function then becomes,
		\begin{equation*}
			J[\{\textbf{u}^{(i)}_s\}_{i=1,s=1}^{i=N,s=S}] = g(\{\textbf{x}^{(i)}_S\}_{i=1}^N) + \Delta t\, \sum_{s=0}^{S-1} L(\{\textbf{x}^{(i)}_s\}_{i=1}^N, \{\textbf{u}^{(i)}_s\}_{i=1}^N).
		\end{equation*}
	
	Then the shooting method starts out by making an initial guess for $\{\textbf{u}^{(i)}_s\}_{i=1,s=1}^{i=N,s=S}$, and then we integrate according to the agent dynamics~\eqref{eq:agent-dynamics}. We then update each $\textbf{u}^{(i)}_s$ by gradient descent on the cost function,
		\begin{equation*}
			\textbf{u}^{(i)}_s \leftarrow \textbf{u}^{(i)}_s - \alpha \nabla_{\textbf{u}^{(i)}_s} J
		\end{equation*}
	where $\alpha$ is the gradient descent step-size. We repeat this iteration until some convergence criterion has reached, e.g. when the controls $\{\textbf{u}^{(i)}_s\}_{i=1,s=1}^{i=N,s=S}$ cease to stop changing very much between iterations.

\section{Numerical Demonstrations}

	Here we provide numerical demonstrations of our method. We first introduce the particular agent dynamics we will use for all our experiments -- the 2D double integrator. Then using the direct shooting method as mentioned in Section~\ref{sec:classical-direct-shooting-for-optimal-control} we examine the cases of having shape constraints in the terminal constraint, then in the running cost. Afterwards, we examine the interplay between shape constraints and obstacle avoidance and congestion/collision minimization. To compute the Earth Mover's Distance, we utilize the Python Optimal Transport (POT) toolbox \cite{flamary2021pot}.
	
\subsection{Agent dynamics and the general cost function}

	In these numerical demonstrations, for each agent we will use the 2D double integrator:
		\begin{equation*}
			\left\{
			\begin{aligned}
				\ddot{x} &= \alpha \\
				\ddot{y} &= \beta \\
				x&(0) = x_0, \;y(0) = y_0, \;\dot{x}(0) = 0,  \;\dot{y}(0) = 0
			\end{aligned}
			\right.
		\end{equation*}
	Of course we then turn this into the first-order system by letting $x_1 = x$, $x_2 = y$, $x_3 = \dot{x}$, $x_4 = \dot{y}$, so if $\textbf{x} = (x_1, x_2, x_3, x_4)^\intercal$, and we let $\textbf{u} = (\alpha, \beta)^\intercal$, so finally we get,
		\begin{equation*}
			\left\{
			\begin{aligned}
				\dot{\textbf{x}}(t) &= \left[\begin{matrix}
					0 & 0 & 1 & 0 \\
					0 & 0 & 0 & 1 \\
					0 & 0 & 0 & 0 \\
					0 & 0 & 0 & 0
				\end{matrix}\right] \textbf{x}(t) + \left[ \begin{matrix}
				0 & 0 \\
				0 & 0 \\
				1 & 0 \\
				0 & 1
			\end{matrix} \right] \textbf{u}(t) \\
			\textbf{x}(0) &= (x_0, y_0, 0, 0)
			\end{aligned}
			\right.
		\end{equation*}
	Our cost functional will generally have the form,
		\begin{equation*}
			\begin{aligned}
			J[\{\textbf{u}^{(i)}\}] &= g_{\text{shape}}(\{\textbf{x}^{(i)}(T)\}_{i=1}^N) + \frac{1}{N}\sum_{j=1}^N \frac{1}{2}\|(x_3(T), x_4(T))\|^2 \\
			& \qquad + \int_0^T \sum_{j=1}^N \frac{1}{2} \|\textbf{u}^{(j)}(s)\|^2 + L_{\text{shape}}(\{\textbf{x}^{(i)}(t)\}_{i=1}^N)\, ds
			\end{aligned}
		\end{equation*}
	So in all cases, we want the agents to have near zero velocity at terminal time, and we are at least minimizing the squared norm of the control in the running cost.

\subsection{Shape constraints at terminal time}
	
	In order to employ shape constraints at terminal time, we have reference points $\{\textbf{z}^{(j)}\}_{j=1}^M$ and letting $\textbf{a}$ be the weights for the agents and $\textbf{b}$ be the weights for the reference points, then we have
		\begin{equation*}
			g_{\text{shape}}(\{\textbf{x}^{(i)}(T)\}_{i=1}^N) = \sum_{ij} \pi^*_{ij} c_{ij}
		\end{equation*}
	for a given cost matrix $C = (c_{ij})$, and optimal transport plan $\pi^*$. Here, we just let the cost matrix be the squared Euclidean distance between agents and reference points, i.e. $c_{ij} = \|\textbf{x}^{(i)}(T) - \textbf{z}^{(j)}\|^2$.
	
\begin{figure}[!htb]
		\centering
		\begin{minipage}{0.5\textwidth}
			\centering
			\includegraphics[width=1\textwidth]{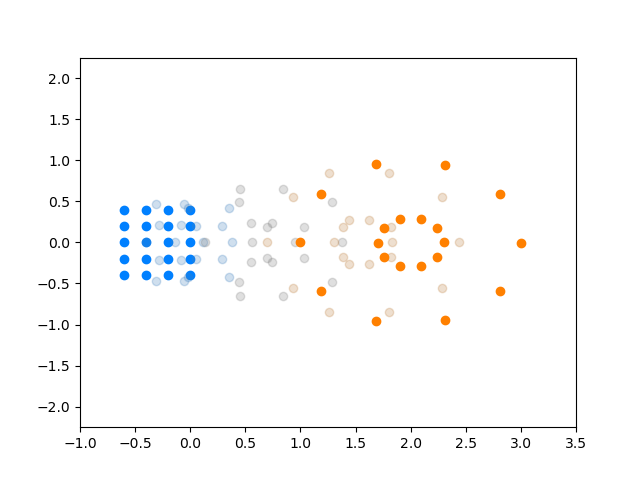}
			\caption{Agents start from a rectangular shape/distribution and move to a ``circle in a circle" shape/distribution. The agents at the starting and ending times are opaque and colored in blue and orange, whereas agents at intermediate times are plotted to be transparent.}
			\label{fig:terminal-circle}
		\end{minipage}%
		\begin{minipage}{0.5\textwidth}
			\centering
			\includegraphics[width=1\textwidth]{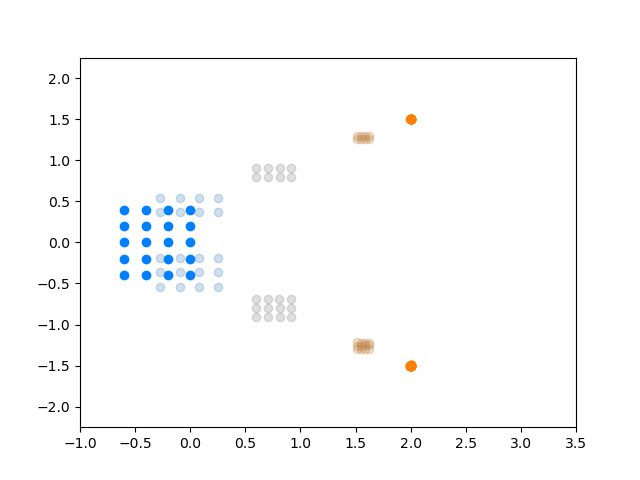}
			\caption{Agents start from a rectangular shape/distribution and their destination is both $(2, 1.5)$ and $(2, -1.5)$. We also enforce that they divide themselves proportionally in a $2/5 - 3/5$ proportion.}
			\label{fig:prop-split}
	\end{minipage}
	\end{figure}

	In Figure \ref{fig:terminal-circle}, we show that the agents are able to move to a ``circle within a circle" shape/distribution. The agents at the starting time and terminal times are opaque and colored in blue and orange respectively. Intermediate times are transparent.

	We can also enforce proportionality constraints on the agents. In Figure \ref{fig:prop-split}, we have the agents move to the destinations $(2, 1.5)$ and $(2, -1.5)$, but enforce that $2/5$ of the agents go to the upper destination, and the rest $3/5$ go to the lower destination. We note that in this case, the number of reference points $\{\textbf{z}^{(j)}\}$ is $2$, and we merely let $\textbf{b} = (\frac{2}{5}, \frac{3}{5})$. This saves on computational cost as compared to the full assignment method as noted in Section \ref{sec:background-emd}.
	
\subsection{Shape constraints in the running cost}

\begin{figure}[!htb]
		\centering
		\begin{minipage}{0.5\textwidth}
			\centering
			\includegraphics[width=1\textwidth]{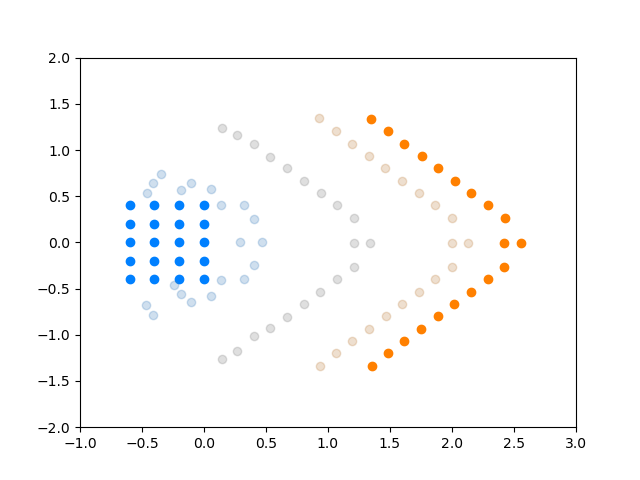}
			\caption{Agents forming a ``Flying V" along the path. This is a demonstration of applying shape constraints in the running cost.}
			\label{fig:running-flyv}
		\end{minipage}%
		\begin{minipage}{0.5\textwidth}
			\centering
			\includegraphics[width=1\textwidth]{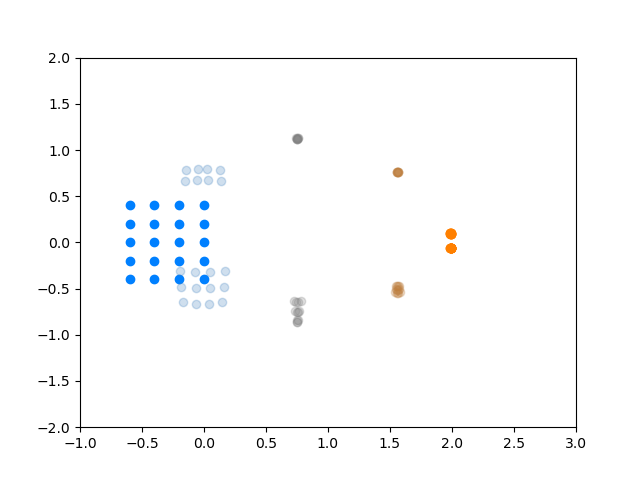}
			\caption{Agents performing a pincer maneuver but also maintaining a $(\frac{2}{5}, \frac{3}{5})$ split along the path.}
			\label{fig:running_pincer}
		\end{minipage}
	\end{figure}

	We can also enforce the agents to adhere to shape constraints along the path fo the agents. In the running cost, we add the term,
		\begin{equation*}
			L_{\text{shape}}(\{\textbf{x}^{(i)}(t)\}_{i=1}^N) = \sum_{ij} \pi^*_{ij}(t) c_{ij}(t)
		\end{equation*}
	In Figure \ref{fig:running-flyv} we make the agents maintain a ``Flying V" shape along the path. In this case, in order to maintain the shape even at terminal time, we also apply the same shape constraint at terminal time, and we loosen the constraint so that it is merely the agent average that should reach the destination of $(2, 0)$.

	We can also enforce proportionality constraints in the running cost as well. In Figure \ref{fig:running_pincer}, we enforce the agents to perform a pincer maneuver, but to also split in a $(\frac{2}{5}, \frac{3}{5})$ proportion. Also, as in Figure \ref{fig:prop-split}, the number of reference points is $2$ again, saving on computational costs.

\subsection{Congestion and Obstacles}

	Here we demonstrate that we can still enforce shape constraints while avoiding collisions and obstacles.
	
\begin{figure}[!htb]
		\centering
		\begin{minipage}{0.5\textwidth}
			\centering
			\includegraphics[width=1\textwidth]{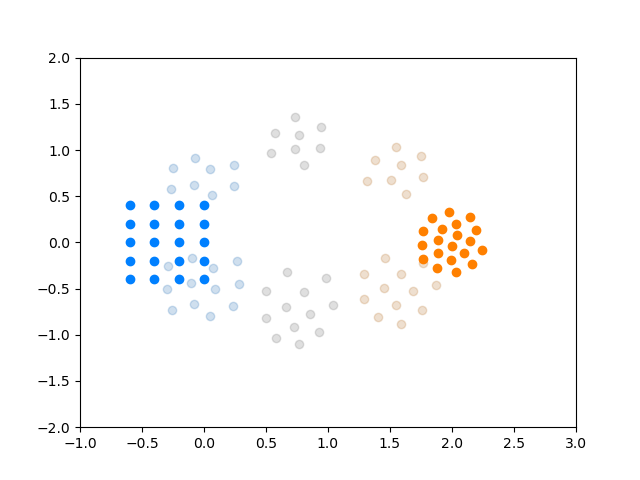}
			\caption{Agents performing a pincer maneuver but now with a congestion/collision penalty, but also maintaining a $(\frac{2}{5}, \frac{3}{5})$ split.}
			\label{fig:running_pincer_cong}
		\end{minipage}%
		\begin{minipage}{0.5\textwidth}
			\centering
			\includegraphics[width=1\textwidth]{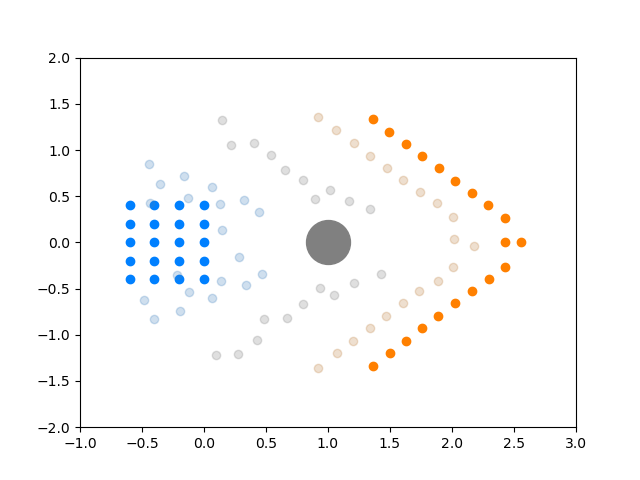}
			\caption{Agents forming a ``Flying V" formation, but now with a congestion/collision penalty, and avoiding an obstacle.}
			\label{fig:running-flyv-cong-obst}
		\end{minipage}
	\end{figure}

	In Figure \ref{fig:running_pincer_cong}, we perform the same pincer maneuver as in Figure \ref{fig:running_pincer}, but now we apply a congestion penalty. This is enforced by using the Gaussian kernel in the running cost:
		\begin{equation*}
			\text{congestion penalty} = \sum_{ij} \text{exp}\left(-\frac{1}{2\sigma^2} \left\| \textbf{x}^{(i)}(t) - \textbf{x}^{(j)}(t) \right\|^2 \right)
		\end{equation*}
	where we chose $\sigma = 0.15$. We see that the agents do indeed avoid colliding.

	And in Figure \ref{fig:running-flyv-cong-obst}, the agents form the same ``Flying V" as in Figure \ref{fig:running-flyv}, but now with congestion/collision penalty, and avoiding an obstacle. We note that compared to the case without an obstacle, the ``tip" of the ``Flying V" differ. As the agents move towards their destination, we can see a dynamic reassigning of roles -- initially the ``tip" of ``Flying V" is the middle agent in the right-furthest column, as can be seen in Figure \ref{fig:running-flyv}. But with an obstacle, the agents dynamically re-assign themselves, so now the role of the ``tip" is a different agent.

\section{Conclusion}
	In this work, we demonstrate that with MASCOT using Optimal Transport and the Earth Mover's Distance, we can apply shape constraints to enforce shape/formation/density constraints on agents. We provide numerical examples and demonstrate that dynamic re-assigning of roles can take place.

\section{Acknowledgments}

	Alex Tong Lin and Stanley J. Osher were supported by the grants: AFOSR MURI FA9550-18-1-0502, and ONR grants: N00014-20-1-2093,  N00014-20-1-2787, and also by DOD-AF-AIR FORCE RESEARCH LABORATORY (AFRL) Sponsor Award Number: 21-EPA-RQ-50:002, and Alex Tong Lin was further supported by Air Fare AWARD 16-EPA-RQ-09.

\bibliography{references}
\bibliographystyle{plain}
\end{document}